\documentclass[twocolumn,superscriptaddress,aps,prl]{revtex4}
\usepackage{graphicx}
\begin{document}

\title{Half-life of the Doubly-magic r-Process Nucleus $^{78}$Ni}

\author{P.T.~Hosmer}
\affiliation{National Superconducting Cyclotron Laboratory, Michigan State 
University, East Lansing, MI 48824, USA}
\affiliation{Dept. of Physics and Astronomy, Michigan State University, East 
Lansing,  MI 48824, USA}

\author{H.~Schatz}
\affiliation{National Superconducting Cyclotron Laboratory, Michigan State 
University, East Lansing, MI 48824, USA}
\affiliation{Dept. of Physics and Astronomy, Michigan State University, East 
Lansing,  MI 48824, USA}
\affiliation{Joint Institute for Nuclear Astrophysics, Michigan State University,
East Lansing, MI 48824, USA}

\author{A.~Aprahamian}
\affiliation{Joint Institute for Nuclear Astrophysics, Michigan State University,
East Lansing, MI 48824, USA}
\affiliation{Dept. of Physics, University of Notre Dame, Notre Dame, 
IN 46556-5670, USA}

\author{O.~Arndt}
\affiliation{Institut f{\"u}r Kernchemie, Universit{\"a}t Mainz, 
Fritz-Strassmann Weg 2, D-55128 Mainz, Germany}

\author{R.R.C.~Clement\footnote{Current affiliation: Lawrence Livermore National 
Laboratory, 7000 East Ave. Livermore, CA 94550, USA}}
\affiliation{National Superconducting Cyclotron Laboratory, Michigan State 
University, East Lansing, MI 48824, USA}

\author{A.~Estrade}
\affiliation{National Superconducting Cyclotron Laboratory, Michigan State 
University, East Lansing, MI 48824, USA}
\affiliation{Dept. of Physics and Astronomy, Michigan State University, East 
Lansing,  MI 48824, USA}

\author{K.-L.~Kratz}
\affiliation{Institut f{\"u}r Kernchemie, Universit{\"a}t Mainz, 
Fritz-Strassmann Weg 2, D-55128 Mainz, Germany}
\affiliation{HGF Virtuelles Institut f{\"u}r Kernstruktur und Nukleare 
Astrophysik (http://www.vistars.de), Mainz, Germany }

\author{S.N.~Liddick}
\author{P.F.~Mantica}
\affiliation{National Superconducting Cyclotron Laboratory, Michigan State 
University, East Lansing, MI 48824, USA}
\affiliation{Dept. of Chemistry, Michigan State University, East Lansing, 
MI 48824, USA}

\author{W.F.~Mueller}
\affiliation{National Superconducting Cyclotron Laboratory, Michigan State 
University, East Lansing, MI 48824, USA}

\author{F.~Montes}
\affiliation{National Superconducting Cyclotron Laboratory, Michigan State 
University, East Lansing, MI 48824, USA}
\affiliation{Dept. of Physics and Astronomy, Michigan State University, 
East Lansing,  MI 48824, USA}

\author{A.C.~Morton\footnote{Current affiliation: TRIUMF, 4004 Wesbrook Mall, 
Vancouver, BC V6T 1R9 Canada}}
\affiliation{National Superconducting Cyclotron Laboratory, Michigan State 
University, East Lansing, MI 48824, USA}

\author{M.~Ouellette}
\author{E.~Pellegrini}
\affiliation{National Superconducting Cyclotron Laboratory, Michigan State 
University, East Lansing, MI 48824, USA}
\affiliation{Dept. of Physics and Astronomy, Michigan State University, East 
Lansing,  MI 48824, USA}

\author{B.~Pfeiffer}
\affiliation{Institut f{\"u}r Kernchemie, Universit{\"a}t Mainz, 
Fritz-Strassmann Weg 2, D-55128 Mainz, Germany}

\author{ P.~Reeder}
\affiliation{Richland, WA 99352, USA}

\author{P.~Santi\footnote{Current affiliation: Los Alamos National Laboratory, 
Safeguards Science and Technology Group (N-1), E540, Los Alamos, 
NM 87544, USA}}
\affiliation{National Superconducting Cyclotron Laboratory, Michigan State 
University, East Lansing, MI 48824, USA}

\author{M.~Steiner}
\author{A.~Stolz}
\affiliation{National Superconducting Cyclotron Laboratory, Michigan State 
University, East Lansing, MI 48824, USA}

\author{B.E.~Tomlin}
\affiliation{National Superconducting Cyclotron Laboratory, Michigan State 
University, East Lansing, MI 48824, USA}
\affiliation{Dept. of Chemistry, Michigan State University, East Lansing, 
MI 48824, USA}

\author{W.B.~Walters}
\affiliation{Dept. of Chemistry and Biochemistry, University of Maryland, 
College Park, MD 20742, USA}

\author{A.~W{\"o}hr}
\affiliation{Dept. of Physics, University of Notre Dame, Notre Dame, 
IN 46556-5670, USA}

\date{\today}

\begin{abstract}

Nuclei with magic numbers serve as important benchmarks in nuclear theory.
In addition, neutron-rich nuclei play an important role in the astrophysical 
rapid neutron-capture process (r-process).$^{78}$Ni is the only doubly-magic 
nucleus that is also an important waiting point in the r-process, and
serves as a major bottleneck in the synthesis of heavier elements. The 
half-life of $^{78}$Ni has been experimentally deduced for the first time 
at the Coupled Cyclotron Facility of the National Superconducting Cyclotron 
Laboratory at Michigan State University, and was found to be 
$110^{+100}_{-60}$ ms. In the same experiment, a first 
half-life was deduced for $^{77}$Ni of $128^{+27}_{-33}$ ms, and more precise 
half-lives were deduced for $^{75}$Ni and $^{76}$Ni of $344^{+20}_{-24}$ ms and 
$238^{+15}_{-18}$ ms respectively.

\end{abstract}

\pacs{}

\keywords{}

\maketitle


Doubly-magic nuclei with completely filled proton and neutron shells are of 
fundamental interest in nuclear physics. The simplified structure of these 
nuclei and their direct neighbors allows one to benchmark key ingredients 
in nuclear structure theories such as single-particle energies and effective 
interactions. Doubly-magic nuclei also serve as cores for shell
model calculations, dramatically truncating the model space, thus rendering
feasible shell model calculations in heavy nuclei. All this is of particular 
importance for nuclei far from stability, where doubly-magic nuclei serve as 
beachheads in the unknown territory of the chart of nuclides \cite{Bro02,DoN98}.

When considering the classic nuclear shell gaps and excluding superheavy 
nuclei, there are only 10 doubly-magic nuclei, and only four of these are 
far from stability: $^{48}$Ni, $^{78}$Ni, $^{100}$Sn, and $^{132}$Sn. Of 
these, $^{48}$Ni and $^{78}$Ni are the most exotic ones, and the last ones 
with experimentally unknown properties. $^{78}$Ni therefore represents a 
unique stepping stone towards the physics of extremely neutron-rich nuclei.
In a pioneering experiment, Engelmann {\it et al.} \cite{Engelmann95} were 
able to identify three $^{78}$Ni events produced by in-flight fission of a 
uranium beam at the Gesellschaft f{\"u}r Schwerionenforschung (GSI), 
demonstrating the existence of this nuclide. We report 
the first measurement of the half-life of $^{78}$Ni at Michigan State University's 
National Superconducting Cyclotron Laboratory (NSCL), demonstrating that 
experiments with $^{78}$Ni are finally feasible. Such a measurement provides 
a first constraint for nuclear models and can serve as a first indicator of 
nuclear properties far from stability (See for example \cite{DKW03}.).

Very neutron-rich nuclei play an important role in the astrophysical 
rapid neutron-capture process (r-process) \cite{Thielemann93,Pfeiffer01}. 
The r-process is responsible for the origin of about half of the heavy 
elements beyond iron in nature, yet its site and exact mechanism are still 
unknown. $^{78}$Ni is the only doubly-magic nucleus that represents an 
important waiting point in the path of the r-process, where the reaction 
sequence halts to wait for the decay of the nucleus \cite{Kratz93}.

One popular astrophysical site for the r-process is the neutrino driven wind 
off a hot, newborn neutron star in a core-collapse supernova 
explosion \cite{WoH92}. In this case the r-process begins around mass 
number $A=90$, with lighter nuclei being produced as less neutron-rich 
species in an $\alpha$-rich freeze-out. For such a scenario $^{78}$Ni would 
not be directly relevant. However, the $\alpha$-rich freezeout fails to 
accurately reproduce the observed abundances for nuclei 
with $A=80-$90 \cite{FRR99}, and the associated r-process does not produce 
sufficient amounts of the heaviest r-process nuclei
around $A=$195 \cite{TWJ94}.

$^{78}$Ni is among the important r-process waiting points in models that 
try to address these issues. Examples include models that assume nonstandard 
neutron star masses \cite{TSK01}, or that are based on a supernova triggered 
by the collapse of an ONeMg core in an intermediate mass star \cite{WTI93}. 
In these models the neutron-capture process begins at lighter nuclei and the 
half-life of $^{78}$Ni becomes a direct input. Together with the other already 
known waiting points, $^{79}$Cu and $^{80}$Zn, the half-life of $^{78}$Ni sets 
the r-process timescale through the $N=50$ bottleneck towards heavier elements, 
and also determines the formation and shape of the associated $A=$ 80 abundance 
peak in the observed r-process element abundances. The $A=80$ mass region has 
recently gained importance in light of new observations of the element 
abundances produced by single (or very few) r-process events as preserved in 
the spectra of old, very metal-poor stars in the Galactic halo. These 
observations point to the possibility of two different 
r-processes \cite{PKO00,TGA04} being responsible for the origin of light 
r-process nuclei below $A < 130$. Only with accurate nuclear data,
especially around $^{78}$Ni - $^{80}$Zn, will it be possible to disentangle
the various contributions from neutron-capture processes in different 
astrophysical sites, and to interpret the data on neutron-capture elements 
expected from the many new metal-poor stars to be identified in ongoing 
surveys \cite{BBC04}.


In this experiment a
secondary beam comprised of a mix of several neutron-rich nuclei 
around $^{78}$Ni was produced by fragmentation of a 140~MeV/nucleon 
$^{86}$Kr$^{34+}$ primary beam on a 376~mg/cm$^2$ Be target at the NSCL 
Coupled Cyclotron Facility. The average primary beam intensity was 15~pnA. 
Fragments were separated by the A1900 fragment separator \cite{Morr03} 
operating with full momentum acceptance. A position sensitive plastic 
scintillator at the dispersive intermediate focus was used to determine the 
momentum of each beam particle at typical rates of $10^{5}$/s. 
A 100.9 mg/cm$^2$ achromatic Al degrader 
was also placed at the intermediate focus of the separator to provide 
increased separation.

Each nucleus in the secondary beam was individually identified in-flight by 
measuring energy loss and time of flight, together with the A1900 momentum 
measurement. The time of flight was measured between two scintillators 
separated by about 40 m: one located at the intermediate image of the A1900 
and the other located inside the experimental vault. The beam was stopped 
in a stack of Si detectors of the NSCL Beta Counting System 
(BCS) \cite{Prisc03}. Energy loss was measured in the first two Si 
detectors, which were separated by a passive Al degrader of variable 
thickness.  The degrader thickness was adjusted to stop the nuclei in a 
985 $\mu$m double-sided Si strip detector (DSSD). The DSSD was segmented into 
40 1 mm strips horizontally on one side, and vertically on the other, 
resulting in 1600 pixels. The beam was continuously implanted into the DSSD, 
which registered the time and position of each ion.  The typical total 
implantation rate for the entire detector was under 0.1 per second.

\begin{figure}
\begin{center}
\includegraphics[scale=0.35]{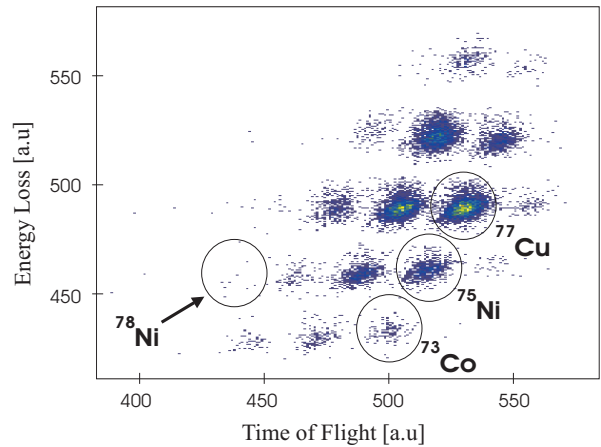}
\caption{Particle identification using energy loss vs. time of flight for a 
subset of the data.}
\label{PID}
\end{center}
\end{figure}

Using the dual-gain capability of the BCS electronics, the DSSD also 
registered the time and position of any $\beta$-decays following the 
implantation of a nucleus. This allowed the correlation of a decay event with 
a previously identified implanted nucleus. Additonal Si detectors in front and 
behind the DSSD were used to veto events from light particles in the secondary 
beam that can be similar to $\beta$-decay events. With this setup, the total 
$\beta$-type event background rate associated with an implanted ion was 
typically less than $3 \times 10^{-2}$/s. Fig.~\ref{PID} shows the particle 
identification using energy loss vs. time of flight. A total of 11 $^{78}$Ni 
events were identified over a total beam time of 104 hours. Using a rough 
estimate for the integrated beam current and a transmission for $^{78}$Ni 
fragments into the experimental vault of 65\% calculated with the Monte-Carlo 
beam transport code MOCADI \cite{Iwasa97}, we obtain a rough estimate of the 
production cross section of 0.02$\pm 0.01$ pb. This is lower than the 
estimated cross section from the EPAX formula of 4 pb \cite{SuB00}.

Decay half-lives were determined using a maximum likelihood analysis,
which has been used before in experiments with low statistics \cite{Bernas90}, 
and in extreme cases with just 6 and 7 $\beta$-decay events \cite{Schneider95}.
For this work, the formalism was modified to account for $\beta$-delayed
neutron emission. The method finds the decay constant that maximizes a 
likelihood function, which is the product of probability densities for three 
decay generations as well as background events, to produce the measured time 
sequence of decay-type events following the implantation of a beam particle.  
The calculation requires knowledge of the $\beta$-detection efficiency, 
background rate, daughter and granddaughter half-lives, including those 
reached by $\beta$-delayed neutron emission, and  branchings for 
$\beta$-delayed neutron emission ($P_n$) for all relevant nuclei in the 
decay chain.

For $^{75}$Ni, $^{76}$Ni,$^{77}$Cu and $^{78}$Cu the statistics were 
sufficient to determine the $\beta$-detection efficiency by comparing fitted 
decay curves with the total number of implanted species of that isotope. The 
resulting efficiencies agree very well and range from 40\% to 43\% with no 
systematic trends in the deviation. For $^{77}$Ni and $^{78}$Ni, an average 
efficiency of (42 $\pm$1)\% was adopted. The background was determined for 
each run (typical duration of 1h) and in each detector pixel by counting all 
decay events that occur outside of a 100~s time window following an 
implantation. Because of the low implantation rate the background is constant 
over the 5 s time window used to correlate decays to an implantation. 
Experimental $P_n$ values as well as daughter and grand-daughter half-lives 
used for the analysis were taken from \cite{nubase03} and \cite{Pfeiffer02} 
when available.  The experimentally unknown $P_n$ values for the Ni isotopes 
were taken from detailed spherical quasi-particle random-phase (QRPA) 
calculations for pure Gamow-Teller (GT) and GT with first-forbidden 
decay \cite{MoR90} and a number of different choices of single-particle 
potentials and mass model predictions. From this study, we derive an average
uncertainty for the calculated $P_n$ values of about a factor of two.

The statistical error of the derived decay half-lives is obtained directly 
from the maximum likelihood analysis. As sources of systematic errors we 
considered uncertainties in the $P_n$ values and daughter or grand-daughter 
half-lives, as well as uncertainties in background rate and detection 
efficiencies. The systematic uncertainties for the half-lives of
$^{78}$Ni, $^{77}$Ni, $^{76}$Ni, $^{75}$Ni are, (in ms)
$^{+33}_{-10}$, $^{+11}_{-7}$, $^{+6}_{-5}$, $^{+8}_{-6}$, respectively.
The main contribution to the systematic errors are uncertainties in the 
detector efficiency, and uncertainties in the parent $P_n$ values. In the case 
of $^{78}$Ni, we also took into account the possibility that one of the events 
is misidentified. Given the very low number of events beyond $^{78}$Ni in the 
particle identification (see Fig.~\ref{PID}) this is a very conservative 
assumption. For $^{78}$Ni this leads to a systematic error of $^{+10}_{-0}$ms.

Systematic and statistical errors are correlated since the shape of
the likelihood function depends on the analysis parameters.
To add systematic and statistical errors we therefore reran
the analysis for all combinations of systematic variations and employed
the lower and upper one-sigma limits of the resulting statistical errors
as the total error budget.

In principle our 
analysis depends somewhat on the unknown feeding and decay branchings of the
known isomeric states in $^{76}$Cu and $^{77}$Zn, which are part of the decay
chains considered here. Assuming decay from 
the isomeric state with a half-life of 1.27 s for $^{76}$Cu would increase the 
$^{76}$Ni half-life by no more than 12 ms and the $^{77}$Ni half-life by no 
more than 5 ms. Assuming population of the 1.05 s isomer for $^{77}$Zn could 
change the half-life of $^{77}$Ni by -8 ms to +13 ms, and the half-life of 
$^{78}$Ni by -10 ms to +15 ms depending on the probability for that state
to $\beta$-decay. These uncertainties are based on extreme assumptions with
no obvious central value. We therefore give them separately and do not include
them in our systematic error bars. 

For the assumption that $^{76}$Cu and $^{77}$Zn $\beta$-decay from the 
ground states with half-lives of 0.641 s and 2.08 s respectively \cite{nubase03},
our final results are $344^{+20}_{-24}$ ms for $^{75}$Ni, 
$238^{+15}_{-18}$ ms for $^{76}$Ni, $128^{+27}_{-33}$ ms for $^{77}$Ni, 
and $110^{+100}_{-60}$ ms for $^{78}$Ni. For $^{77}$Cu and $^{78}$Cu, we obtain 
$450^{+13}_{-21}$ ms and $323^{+11}_{-19}$ ms, in excellent agreement with 
previous work (469$\pm$8 ms and 342$\pm$11 ms) \cite{KGM94}.

\begin{figure}
\begin{center}
\includegraphics[scale=0.35]{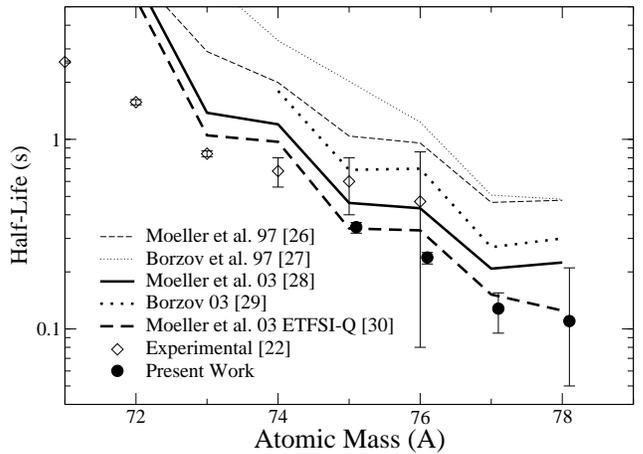}
\caption{Experimental Ni half-lives from this and previous work \cite{nubase03} 
compared to different theoretical calculations.}
\label{theorycompare}
\end{center}
\end{figure}

In Fig.~\ref{theorycompare} our new experimental half-lives are compared with 
various theoretical predictions. Often employed in r-process model calculations 
are the global QRPA calculations of M\"oller {\it et al.} 1997 \cite{MNK97} 
or Borzov {\it et al.} 1997 \cite{BoGoPe97}, the latter being limited
to spherical nuclei. Our results show that the trend of these models to 
over-predict Ni half-lives by factors of 3-4 already observed for the more 
stable isotopes persists into the path of the r-process at $^{78}$Ni.
The recent versions of both models \cite{MPK03,Bor03} besides other 
improvements now also include first-forbidden transitions.  They clearly lead 
to better though still somewhat large half-life predictions. 
Fig.~\ref{theorycompare} also shows results from calculations with the same
model as M\"oller {\it et al.} 2003 \cite{MPK03} but using a mass model that 
includes a quenching of shell gaps far from stability 
[extended Thomas-Fermi approach + Strutinsky Integral, with shell 
quenching (ETFSI-Q) \cite{Pearson96}]. These calculations give the best 
agreement with experimental data among the global models.

\begin{figure}
\begin{center}
\includegraphics[scale=0.35]{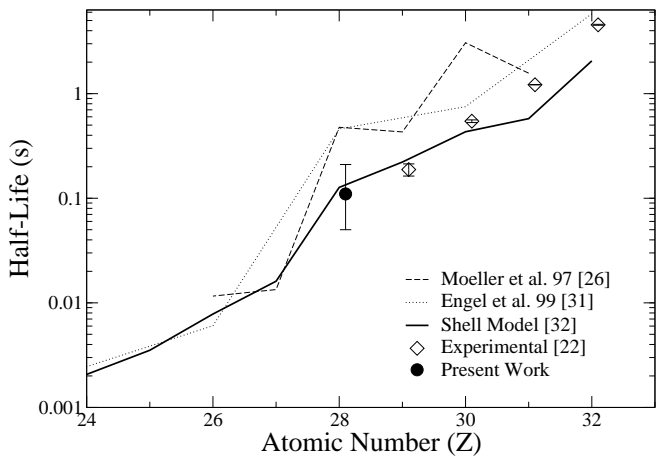}
\caption{Theoretical half-life calculations for $N=50$ compared to experimental 
data from this and previous work \cite{nubase03}. Engel {\it et al.} only gives
values for even $Z$ nuclei.}
\label{n_theorycompare}
\end{center}
\end{figure}

In order to better understand the nuclear structure in this mass region and to 
benchmark global models beyond the range of experimental data it is important 
to test the more sophisticated microscopic calculations, which have been
performed for a limited set of singly- and doubly-magic heavy nuclei
(see Fig.~\ref{n_theorycompare}). The self-consistent QRPA approach
\cite{EBD99} agrees with the shell-model calculation \cite{LaM03} and the 
experimental data for most nuclei, but predicts a $^{78}$Ni half-life that 
even exceeds the experimental $^{80}$Zn half-life. Our measurement 
clearly favors a much lower $^{78}$Ni half-life. On the other hand,
the shell model results are in good agreement with experimental data. 
Of course this does not necessarily mean that the shell-model description of 
this mass region is entirely correct. For example, deviations in excitation 
energies, transition strengths, and decay Q-value can in principle compensate 
each other. More experiments including detailed spectroscopy as they might 
become possible at future facilities will be needed to clarify this.

In summary, we present the first results for the half-life of $^{78}$Ni and 
other neutron-rich Ni isotopes. With these results, experimental half-lives 
are available for all but one ($^{48}$Ni) classical doubly-magic nuclei. Also, 
the half-lives of all important $N=50$ waiting points in the r-process are now 
known experimentally. This will make r-process model predictions of the 
nucleosynthesis around $A=80$ more reliable and comparison with observational 
data more meaningful. It will also put the overall delay that the $N=50$
mass region imposes on the r-process flow towards heavier elements on a
more solid experimental basis. In this respect the half-life of $^{78}$Ni is
of special importance as during the initial stages of the r-process
when the heavier nuclei are synthesized the r-process path
passes through $^{78}$Ni and $^{79}$Cu rather than through the more 
stable $N=50$ nuclei \cite{Wan04}. The delay timescale for the buildup of 
heavy elements beyond $N=50$ is therefore set by the sum of the lifetimes 
of $^{78}$Ni and $^{79}$Cu. Our experimental data clearly favor the short 
timescale of 450~ms obtained with the prediction of Langanke and 
Martinez-Pinedo \cite{LaM03} over the much longer delays of 960~ms predicted 
for example by M{\"oller} {\it et al.} \cite{MNK97} leading to an acceleration 
of the r-process. This is in line with recent improvements in theoretical
$\beta$-decay half-life predictions along the entire r-process path
that also tend to result in shorter half-lives thereby
speeding up the r-process \cite{MPK03}.
Detailed r-process model calculations with the new experimental data
are beyond the scope of this paper,
but will be presented in a forthcoming study.

This work has been supported by NSF grants PHY 01-10253 and PHY 02-16783
(Joint Institute for Nuclear Astrophysics), 
DFG grant KR 806/13-1, and
HGF grant VH-VI-061.
H. S. is supported by the Alfred P. Sloan Foundation.


\begin{thebibliography}{9}

\bibitem{Bro02}
B. A. Brown, Nucl. Phys. A{\bf 704}, 11c (2002).

\bibitem{DoN98}
J. Dobaczewski and W. Nazarewicz, Phil. Trans. Math. Phys. Eng. Sci. 
{\bf 356}, 2007 (1998).

\bibitem{Engelmann95}
Ch. Engelmann {\it et. al.}, Z. Phys. A {\bf 352} 351 (1995). 

\bibitem{DKW03}
I. Dillmann {\it et. al.}, Phys. Rev. Lett. {\bf 91}, 162503 (2003).

\bibitem{Thielemann93}
F.-K. Thielemann {\it et al.}, Phys. Rep. {\bf  227}, 269 (1993).

\bibitem{Pfeiffer01}
B. Pfeiffer {\it et. al.}, Nucl. Phys. A {\bf 693}, 282 (2001).

\bibitem{Kratz93}
K.-L. Kratz {\it et al.}, Ap. J. {\bf 403}, 216 (1993).

\bibitem{WoH92}
S. E. Woosley and R. D. Hoffman, Ap. J. {\bf 395}, 202 (1992).

\bibitem{FRR99}
C. Freiburghaus {\it et al.}, Ap. J. {\bf 525}, L121 (1999).

\bibitem{TWJ94}
K. Takahashi, J. Witti, and H.-T. Janka, Astron. Astrophys. {\bf 286}, 
857 (1994).

\bibitem{TSK01}
T. Terasawa {\it et al.}, Ap. J. {\bf 562}, 470 (2001).

\bibitem{WTI93}
S. Wanajo {\it et al.}, Ap. J. {\bf 593}, 968 (2003).

\bibitem{PKO00}
B. Pfeiffer {\it et al.}, {\it "The First Stars". Proceedings of the MPA/ESO
Workshop held at Garching, Germany, 4-6 August 1999}, edited by A. Weiss, T. G. Abel,
and V. Hill (Springer, Berlin, 2000), p.148.

\bibitem{TGA04}
C. Travaglio  {\it et al.}, Ap. J. {\bf 601}, 864 (2004).

\bibitem{BBC04}
T.C. Beers, P. S. Barklem, N. Christlieb and V. Hill, astro-ph/0408381
to be published in Nucl. Phys. A.

\bibitem{Morr03}
D.J.Morrisey {\it et al.}, Nucl. Instrum. Methods B {\bf 204}, 90 (2003).

\bibitem{Prisc03}
J.I.Prisciandaro {\it et al.}, Nucl. Instrum. Methods A {\bf 505}, 140 (2003).

\bibitem{Iwasa97}
N. Iwasa {\it et al.}, Nucl. Instrum. Methods B {\bf 126}, 284 (1997).

\bibitem{SuB00}
K. S{\"u}mmerer and B. Blank, Phys. Rev. C {\bf 61}, 034607 (2000).

\bibitem{Bernas90}
M.Bernas {\it et al.}, Z. Phys. A {\bf 336} 41 (1990).

\bibitem{Schneider95}
R.Schneider {\it et al.}, Nucl. Phys. A{\bf 588}, c191 (1995).

\bibitem{nubase03}
G. Audi {\it et al.} Nucl. Phys. A{\bf 729}, 3 (2003).

\bibitem{Pfeiffer02}
B.Pfeiffer {\it et al.}, Prog. Nucl. Energy {\bf 41}, 39 (2002).

\bibitem{MoR90} 
P. M\"oller and Randrup, J., Nucl. Phys. A {\bf 514}, 1 (1990).

\bibitem{KGM94}
K.-L. Kratz {\it et al.}, Z. Phys. A{\bf 340}, 419 (1991).

\bibitem{MNK97}
P. M{\"o}ller, J. R. Nix, and K.-L. Kratz, Atomic Data and Nucl. Data Tab. 
{\bf 66}, 131 (1997).

\bibitem{BoGoPe97}
I. N. Borzov, S. Goriely, and J.M. Pearson, Nucl. Phys. A{\bf 621}, 307c 
(1997).

\bibitem{MPK03}
P. M{\"o}ller, B. Pfeiffer, and K.-L. Kratz, Phys. Rev. C {\bf 67}, 055802 
(2003).

\bibitem{Bor03}
I. N. Borzov, Phys. Rev. C {\bf 67}, 025802 (2003).

\bibitem{Pearson96}
J. M. Pearson {\it et al.} Phys. Lett. B {\bf 387}, 455 (1996).

\bibitem{EBD99}
J. Engel {\it et al.} Phys. Rev. C {\bf 60}, 014302 (1999).

\bibitem{LaM03}
K. Langanke and G. Martinez-Pinedo Rev. Mod. Phys. {\bf 75}, 819 (2003).

\bibitem{Wan04}
S. Wanajo, r-process movie, at 
http://www.ph.sophia.ac.jp/~shinya/research/research.html

\end{thebibliography}
\end{document}